\documentclass[prx,aps,floatfix,longbibliography,superscriptaddress,twocolumn,final,notitlepage]{revtex4-2}
\RequirePackage{amsmath}
\RequirePackage{hyperref} 
\RequirePackage{amssymb}
\RequirePackage{graphicx}
\RequirePackage{bbm}
\RequirePackage{xcolor}
\usepackage{lipsum}
\usepackage{comment}

\usepackage{pgf}

\setcitestyle{round}

\usepackage{mathtools}
\usepackage{caption}
\usepackage{subcaption}
\usepackage{amsthm}
\usepackage{graphicx}
\usepackage{duckuments}

\newcommand{\be}{\begin{equation}}
\newcommand{\ee}{\end{equation}}
\newcommand{\bea}{\begin{eqnarray}}
\newcommand{\eea}{\end{eqnarray}}

\newtheorem*{definition*}{Definition}

\usepackage{amsmath,fixmath}
\usepackage{setspace}

\usepackage[noline, linesnumbered]{algorithm2e}
\RestyleAlgo{ruled}
\SetKwComment{Comment}{> }{}
\SetKwInput{KwInput}{Input}
\SetKwInput{KwYield}{Yield}

\usepackage{float}
\usepackage[nameinlink,capitalize]{cleveref} 

\usepackage{booktabs}
\usepackage{adjustbox}
\definecolor{shadecolor}{gray}{0.9}
\usepackage{tabu}

\usepackage[normalem]{ulem} 
\usepackage{soul}

\usepackage{enumitem,amssymb}
\newlist{todolist}{itemize}{2}
\setlist[todolist]{label=$\square$}
\newlist{todolist_done}{itemize}{2}
\setlist[todolist_done]{label=$\blacksquare$}
\usepackage{pifont}
\begin{document}

\title[
    Hypergraphx: a library for higher-order network analysis
    ]{
    Hypergraphx: a library for higher-order network analysis
}

\author{Quintino Francesco Lotito}
\email{quintino.lotito@unitn.it}
\affiliation{Department of Information Engineering and Computer Science, University of Trento, via Sommarive 9, 38123 Trento, Italy}

\author{Martina Contisciani}
	\affiliation{Max Planck Institute for Intelligent Systems, Cyber Valley, 72076 Tübingen, Germany}
\author{Caterina De Bacco}
	\affiliation{Max Planck Institute for Intelligent Systems, Cyber Valley, 72076 Tübingen, Germany}
\author{Leonardo Di Gaetano}
	\affiliation{Department of Network and Data Science, Central European University, 1100 Vienna, Austria}
\author{Luca Gallo}
	\affiliation{Department of Network and Data Science, Central European University, 1100 Vienna, Austria}
\author{Alberto Montresor}
\affiliation{Department of Information Engineering and Computer Science, University of Trento, via Sommarive 9, 38123 Trento, Italy}
\author{Federico Musciotto}
\affiliation{Dipartimento di Fisica e Chimica Emilio Segr\`e, Universit\`a di Palermo, Viale delle Scienze, Ed. 18, I-90128, Palermo, Italy}
\author{Nicol\`o Ruggeri}
	\affiliation{Max Planck Institute for Intelligent Systems, Cyber Valley, 72076 Tübingen, Germany}
	\affiliation{Department of Computer Science,  ETH,  8004 Z\"urich, Switzerland}
\author{Federico Battiston}
	\email{battistonf@ceu.edu}
	\affiliation{Department of Network and Data Science, Central European University, 1100 Vienna, Austria}

\begin{abstract}
From social to biological systems, many real-world systems are characterized by higher-order, non-dyadic interactions. Such systems are conveniently described by hypergraphs, where hyperedges encode interactions among an arbitrary number of units. Here, we present an open-source python library, hypergraphx (HGX), providing a comprehensive collection of algorithms and functions for the analysis of higher-order networks. These include different ways to convert data across distinct higher-order representations, a large variety of measures of higher-order organization at the local and the mesoscale, statistical filters to sparsify higher-order data, a wide array of static and dynamic generative models, and an implementation of different dynamical processes with higher-order interactions. 
Our computational framework is general, and allows to analyse hypergraphs with weighted, directed, signed, temporal and multiplex group interactions.
We provide visual insights on higher-order data through a variety of different visualization tools.
We accompany our code with an extended higher-order data repository, and demonstrate the ability of HGX to analyse real-world systems through a systematic analysis of a social network with higher-order interactions. The library is conceived as an evolving, community-based effort, which will further extend its functionalities over the years.
Our software is available at \href{https://github.com/HGX-Team/hypergraphx}{\url{https://github.com/HGX-Team/hypergraphx}}.
\end{abstract}

\maketitle

\section*{Introduction}
\label{sec:intro}

In the last few decades, networks have emerged as the natural tool to model a wide variety of natural, social and man-made systems. 
Networks, collections of nodes and links connecting pairs of them, are able to capture dyadic interactions only. 
However, in many real-world systems units interact in groups of three or more~\cite{battiston2020networks, battiston2021physics, bianconi2021higher, battiston2022higher}. 
Systems with non-dyadic interactions are ubiquitous, with examples ranging from cellular networks~\cite{klamt2009hypergraphs}, drug recombination~\cite{zimmer2016prediction}, structural and functional brain networks~\cite{petri2014homological,giusti2016two, santoro2023higher},human~\cite{cencetti2021temporal} and animal~\cite{musciotto2022beyond} face-to-face interactions, and collaboration networks~\cite{patania2017shape}.
These higher-order interactions can be naturally described by alternative mathematical structures such as hypergraphs~\cite{battiston2020networks,berge1973graphs}, where hyperedges connect groups of nodes of arbitrary size.

In the last 25 years, advances in technology have generated an unprecedented amount of relational data across a variety of domains. 
Broadening the scopes of the first pioneering contributions to the field of network science~\cite{wasserman1994social,watts1998collective,barabasi1999emergence}, these allowed to develop new data-informed frameworks to investigate biological, technological and social systems.
In parallel with theoretical and methodological progresses, a crucial role in advancing network science has been played by the development of efficient algorithms and computational tools to analyze networked data. 
Nowadays, widely used, community-based software such as NetworkX~\cite{hagberg2008exploring} and igraph~\cite{csardi2006igraph}, and individual efforts such as graph-tool~\cite{peixoto_graph-tool_2014} -- just to mention a few -- have enabled thousands of researchers to perform multi-faceted, large-scale network analysis of relational data.
However, despite some early contributions~\cite{antelmi2020analyzing,marchette2021hyperg,hnx2021hnx,diaz2022hypergraphs,badie2022reticula}, in particular XGI~\cite{landry2022xgi}, the recent explosion of interest in higher-order systems has not yet been matched by the development of comprehensive computational frameworks.

Here, we bridge this gap by presenting hypergraphx (HGX), a multi-purpose, open-source python library for the analysis of networked systems with higher-order interactions.
The library is conceived by researchers with several years of experience and direct contributions to the field of higher-order interactions.
Developed by a diverse multidisciplinary team with complementary skills and expertise, HGX aims to provide, as a single source, a comprehensive suite of tools and algorithms for constructing, storing, analysing and visualizing systems with higher-order interactions. 
These include different ways to convert data across distinct higher-order representations, a large variety of measures of higher-order organization at the local and the mesoscale, statistical filters to sparsify higher-order data, a wide array of static and dynamic generative models, an implementation of different dynamical processes, from epidemics to diffusion and synchronization, with higher-order interactions, and more. 
Our computational framework is general, and allows to analyse hypergraphs with weighted, directed, signed, temporal and multiplex group interactions. 
Beyond experts in the field, we hope that our library will make higher-order network analysis accessible to everyone interested in exploring the higher-order dimension of relational data.

\section*{Tools}
\label{sec:methods}

Here, we discuss the main functionalities provided by HGX. The different tools of our library are illustrated online through detailed, user-friendly tutorials.
The library is conceived as an evolving, community-based effort, which will further extend its functionalities over the years.

\paragraph*{Representations.}
Hypergraphs represent the most general and flexible framework to encode systems with higher-order interactions~\cite{berge1973graphs, battiston2020networks}. However, specific research questions or data features might benefit from alternative higher-order frameworks. We provide functions to easily and efficiently convert higher-order data usually represented as hypergraphs into different representations~\cite{battiston2020networks, torres2021why} such as bipartite networks, maximal simplicial complexes, higher-order line graphs, dual hypergraphs and clique-expansion graphs. 

\paragraph*{Basic node and hyperedge statistics.}
Our library provides simple tools characterizing basic node and hyperedge statistics. These include measures of hyperdegree distributions, both aggregated or separated by order of interactions, as well as measures of correlations among them. We include functions to compute hyperdegree-hyperdegree assortativity, both within and across orders. We provide simple tools to compute hyperedge size distribution in the whole system, as well as at the level of individual nodes.

\paragraph*{Centrality measures.}
Centrality scores are a key tool in network analysis, and allow to quantify the importance or influence of different nodes within a system~\cite{wasserman1994social}. Nodes with high centrality usually have a high number of links, are strategically connected to other influential nodes, or are characterized by both such features. Our library provides a variety of higher-order centrality measures, where interactions in different group sizes are taken into account. These include centrality measures based on node participation in different subhypergraphs~\cite{estrada2006subgraph} and different centrality scores based on spectral approaches~\cite{benson2019three}. We also implement measures of hyperedge centrality based on shortest paths and betweenness flows~\cite{aksoy2020hypernetwork}.

\paragraph*{Motifs.}
Motifs are small recurring patterns of subgraphs that are overrepresented in a network~\cite{milo2002network}. Motif analysis has established itself as a fundamental tool in network science to describe networked systems at their microscale, identifying their structural and functional building blocks~\cite{milo2004superfamilies}. 
We provide an implementation for higher-order motif analysis, extracting overabundant subgraphs of nodes connected by higher-order interactions, as originally defined in Ref.~\cite{lotito2022higher}. Given their widespread applications and expected use on large-scale real-world datasets, we also provide an approximated algorithm for higher-order motif analysis based on hyperedge sampling, able to speed up computations by orders of magnitudes with only a minimal compromise in accuracy~\cite{lotito2022exact}. 

\paragraph*{Mesoscale structures.}
One of the most relevant features of graphs representing real-world systems is community structure~\cite{fortunato2010}. A variety of approaches for community detection on graphs show how these partitions provide meaningful insights into the fundamental patterns underlying node interactions. Recently, methods for defining the mesoscale structure of higher-order networks have been explored. Here, we provide an implementation of a spectral method which recovers hard communities via hypergraph cut optimization~\cite{zhou2006learning}. We also implement different generative models able to extract overlapping communities and jointly infer hyperedges~\cite{contisciani2022inference}, allowing to capture a variety of mesoscale organizations, including both disassortative and assortative community structure~\cite{ruggeri2022inference}. We provide a method able to extract hyperlink communities, where interactions, and not system units, are clustered across different hypergraph modules~\cite{lotito2023hyperlink}. Finally, we provide a method to extract the core-periphery organization of higher-order systems, capturing a group of central and tightly connected nodes in hypergraphs governing the overall system behaviour, inspired by Ref.~\cite{tudisco2023core}.

\paragraph*{Filters.}
Many real-world systems are characterized by an abundance of noisy and redundant interactions, resulting in overly densely connected networks. 
Different filtering techniques have been developed to identify the most informative links by adopting an approach based on statistical validation, 
where the statistical significance of interactions of the real system is evaluated by comparing them with an ensemble of random replicas that preserve some individual features (like degree or strength)~\cite{micciche2019primer}. 
Our library provides a variety of different tools to filter systems with higher-order interactions. These include extracting statistically validated hypergraphs, which are a collection of hyperlinks that are over-expressed representing nodes that are significantly interacting in the same exact group of fixed size~\cite{musciotto2021detecting}, and identifying significant maximally interacting sets, which represent the largest groups of nodes that interact significantly, captured by combining interactions of different orders~\cite{musciotto2022identifying}.

\paragraph*{Generative models.}
The ability to produce synthetic data with different topological characteristics has proven crucial for a variety of tasks, from algorithms benchmarking to the study and testing of non-trivial network statistics~\cite{newman2004finding,lancichinetti2008benchmark}. 
In our library, we offer ready-to-use implementations for various synthetic hypergraph samplers.
We provide functions to build generalised Erd\"os-R\'enyi models, both for uniform (all interactions have the same order) and non-uniform (different orders of interactions) hypergraphs. We implement scale-free random hypergraph models with the possibility of tuning the correlation between the degree sequence among different orders. We also include
a variety of randomization tools and a
configuration model for hypergraphs, where samples are produced respecting given node degree and hyperedge size sequences~\cite{chodrow2020configuration}. Based on a similar mechanism, we implement also a more complex sampler which allows to specify hard and soft community assignments for nodes, and arbitrary community structure, such as assortative and disassortative~\cite{ruggeri2022principled}. 
Finally, we provide a higher-order activity-driven model with group interactions that change in time~\cite{petri2018simplicial} and compute the associated percolation threshold.

\paragraph*{Dynamical processes.}
The structural properties of complex networks shape the dynamical process occurring on top of them~\cite{barrat2008dynamical}. 
Recent works have revealed that higher-order interactions significantly impact various dynamical processes, including percolation~\cite{coutinho2020covering}, diffusion~\cite{schaub2020random,carletti2020random}, pattern formation~\cite{carletti2020dynamical,muolo2023turing}, synchronization~\cite{millan2020explosive,skardal2020higher,lucas2020multiorder,gambuzza2021stability,zhang2023higher}, contagion~\cite{iacopini2019simplicial,de2020social,st2021universal}, and evolutionary games~\cite{alvarez2021evolutionary,civilini2021evolutionary, civilini2023explosive}
We provide functions to investigate several of these processes. 
These include tools to study synchronization with higher-order interactions, from the analysis of the multiorder Laplacian matrix for kuramoto dynamics ~\cite{lucas2020multiorder}, to the implementation of the Master Stability Function approach for synchronization stability~\cite{gambuzza2021stability,gallo2022synchronization}.
We also provide an algorithm to simulate simplicial social contagion~\cite{iacopini2019simplicial}, and analytical and numerical tools to investigate random walks on hypergraphs~\cite{carletti2020random}. 

\paragraph*{Weighted, directed, signed, temporal and multiplex hypergraphs.} Our library is highly flexible. It allows to store and analyze hypergraphs with a rich set of features associated with hyperedges, including interactions of different intensity, directions, sign, that vary in time or belong to different layers of a multiplex system.

\paragraph*{Visualization.}
The adoption of higher-order networks is rapidly increasing, and the development of standard tools to visualize them is still in progress. Our library provides different visualization tools to gain visual insights into the higher-order organization of real-world systems. We provide tools to plot systems with higher-order interactions, where hyperedges of arbitrary size encode relationships among an arbitrary number of nodes.
Due to the rapid combinatorial increase in the number of possible higher-order interactions and their overlaps, such a direct approach is particularly suited for systems with a moderate number of nodes, while such a visualization might not be effective in other cases. Therefore, we provide alternative solutions that may assist the practitioner in a variety of cases, such as relational data with a large number of nodes or large hyperedges. For instance, we give the option to plot the bipartite projection of a hypergraph where the two sets of nodes represent respectively the original system units and the hyperedges in which they take part. We can also plot the hypergraph clique projection, which results in a simple graph where each hyperedge of size $s$ is decomposed into a clique of $\frac{s(s-1)}{2}$ unordered pairwise interactions. Additionally, we implement a multilayer representation of the hypergraph where each layer encodes interactions of a given size, and two nodes are connected in layer $s$ only if they interact in the hypergraph through a hyperedge of size $s$. Finally, we offer a novel way of visualizing hypergraphs, where the hypergraph is represented as a graph whose nodes are pie charts. These pie charts indicate the proportion of interaction sizes for each node, and two nodes are connected when they have significant interactions across multiple orders.

\section*{Data}
Here, we present the dataset repository accompanying our library.
Such a repository is intended to provide an initial core of higher-order relational data, that we aim to expand over the next few years.
We illustrate the functionalities of HGX by performing different higher-order analyses for one of these datasets.

\begin{figure*}
    \centering
    \includegraphics[width=\textwidth]{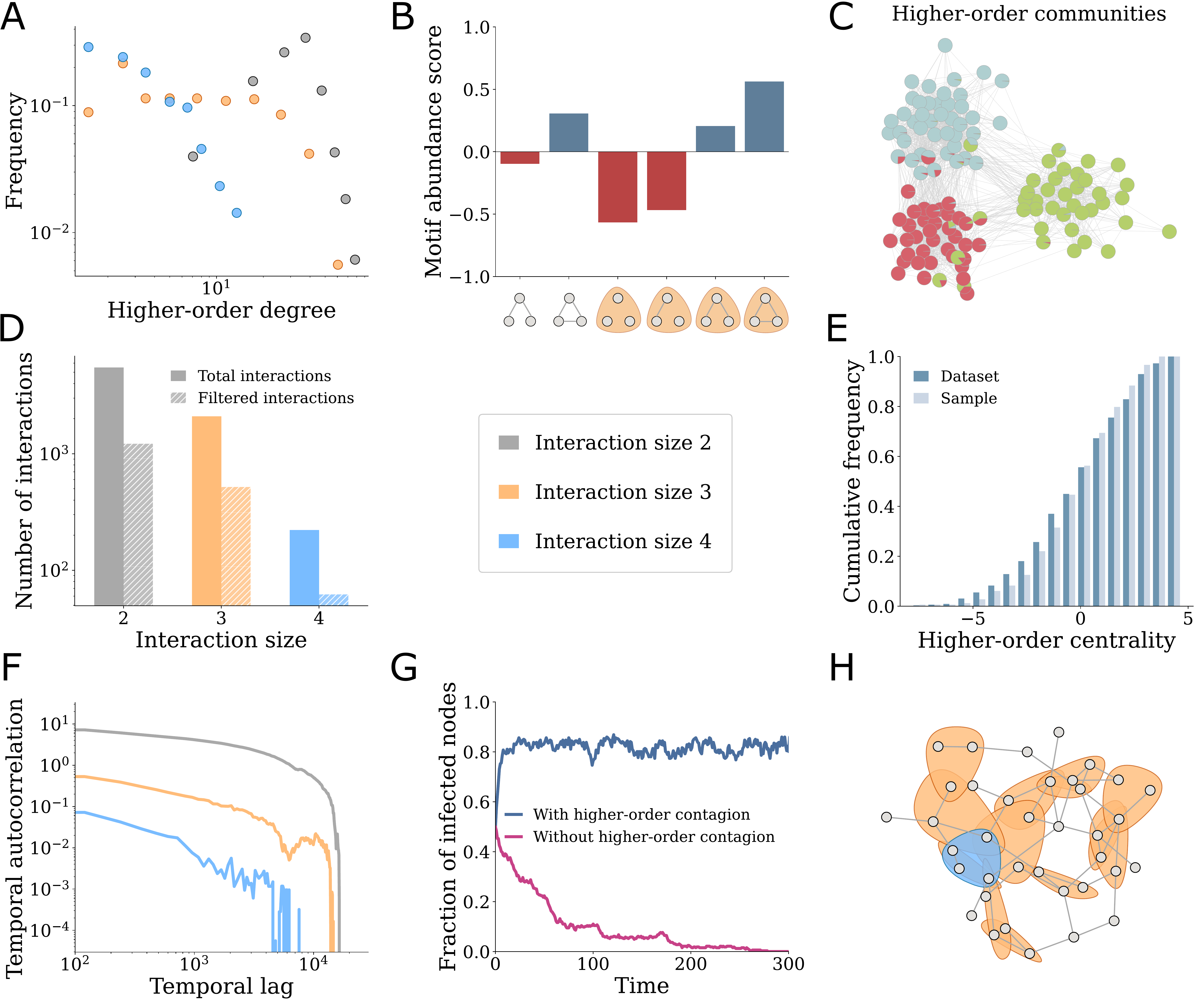}
    \caption{\textbf{Higher-order analysis of social interactions.} We illustrate different functionalities of HGX on a dataset of face-to-face group interactions in a school from the SocioPattern collaboration~\cite{mastrandrea2015contact}. (A) Higher-order degree distributions for different interaction sizes. (B) Higher-order motif analysis. (C) Higher-order overlapping community detection, and comparison with node metadata (we plot a subset of three classes). (D) Statistics of original and filtered higher-order social interactions. (E) Higher-order centrality measure in the dataset, and in sample obtained from a higher-order generative model. (F) Temporal autocorrelation for different sizes. (G) Fraction of infected nodes over time for a spreading process with or without higher-order infections. (H) Direct hypergraph visualization of social interactions (we plot a subset of one class, considering only statistically significant interactions).}
    \label{fig:fig-1}
\end{figure*}

\paragraph*{Higher-order data repository.}
The availability of data plays a fundamental role in developing theoretical frameworks and computational tools across different scientific domains and applications.
The recent explosion of higher-order relational data has led to novel methodologies to study higher-order systems, which in turn require extensive datasets to be tested and validated. 
A few of these data are inherently higher-order. 
Several others, instead, have originally been investigated with pairwise approaches, but have recently been re-explored under the new lens of higher-order network analysis. 
This motivates us to accompany our library with an easily accessible and well-curated data repository, functioning as a unifying source of datasets for the analysis of higher-order systems.
We provide a collection of datasets for higher-order systems across different domains, including ecological (animal proximity~\cite{gelardi2020measuring}), social (human face-to-face interactions~\cite{stehle2011high, vanhems2013estimating, mastrandrea2015contact, genois2015data, genois2018can}, co-authorships~\cite{sinha2015MAG, benson2018simplicial, pacs2021data}, votes~\cite{justice2019database}), technological (e-mails~\cite{leskovec2007graph, yin2017local, benson2018simplicial}) and biological (gene-disease~\cite{bauer2011gene-disease} and drug~\cite{benson2018simplicial} associations) systems. 
Some of these datasets record metadata characterizing the system units (e.g., whether an individual in a hospital is a patient or a doctor) and the interactions among them (e.g., the scientific domain of a research paper involving a group of authors). 
Also, they store information about the structural features of group interactions, which can be nonreciprocal, multi-relational and time-varying. 
Datasets can be loaded to explicitly highlight some of these characteristics.
Indeed, our library allows to apply filters in the data loading process, for example by selecting specific sets of nodes with regard to some metadata restriction, or by extracting group interactions limited to a given size, type or time interval.
In the next years, we plan to continuously expand the data repository, and to add further filtering options to the data loading functions. 

\paragraph*{Analysing real-world higher-order systems: a guided tour.}
To illustrate the power of HGX in loading, manipulating, analysing and visualizing real-world systems with group interactions, in Figure~\ref{fig:fig-1} we present an illustrative analysis of a dataset from the SocioPattern collaboration encoding face-to-face social interactions in a high school~\cite{mastrandrea2015contact}. This dataset has been widely investigated in the literature on higher-order interactions~\cite{benson2018simplicial, iacopini2019simplicial, lotito2022higher, contisciani2022inference, ruggeri2022principled}, and records the activity of $327$ students, divided into nine different high school classes. Our analysis focuses in particular on interactions among $2$, $3$ and $4$ individuals, as statistics is limited for larger groups.

In (A) we show the different higher-order degree distributions. The largest degrees are obtained for pairwise interactions, and, in general, the curves show different profiles. Higher-order degree distributions display different correlations across different orders (Pearson's correlation coefficient $\rho$, $\rho^{2,3}=0.74$, $\rho^{2,4}=0.46$, $\rho^{3,4}=0.72$). To characterize such a higher-order system at the microscale, in (B) we perform higher-order motif analysis as introduced in Ref.~\cite{lotito2022higher}. We consider subhypergraphs of three nodes and capture over- (positive abundance score greater) and under- (negative) represented motifs in the data, as compared to a randomized higher-order configuration model~\cite{chodrow2020configuration}. Local structures with group interactions supported by pairwise links are found to be particularly relevant. In (C) we describe the mesoscale structure of the system, by extracting overlapping communities with the method of Ref.~\cite{contisciani2022inference}. For simplicity, we consider a subset of three classes and plot pairwise interactions only. Nodes are represented as pie-charts, colored proportionally to the higher-order communities they belong to.
In general, the inferred modules are well aligned with node metadata, with most students largely interacting within the community associated with their class. In (D), we show statistics for the interactions in the dataset. We see an inverse trend between the number of interactions and group size. We also plot statistics for a filtered system, where we have considered statistically validated hypergraphs~\cite{musciotto2021detecting}, removing redundant hyperedges and identifying the most informative group interactions. We continue by showcasing the ability of the model introduced in Ref.~\cite{ruggeri2022principled} to generate hypergraphs which are similar to the original dataset. To validate such a statement, in (E) we plot the distribution of (a rescaled version of) higher-order centrality measure~\cite{estrada2006subgraph} both in the real and sampled hypergraphs, showing good agreement between the two. To further illustrate the flexibility of our computational framework, we then consider the temporal dimension of higher-order interactions. In particular, in (F) we  the temporal autocorrelation for different interaction sizes, one of the measures introduced to characterize the temporal evolution of higher-order systems in Ref.~\cite{gallo2023higher}. Results show the existence of long-range correlations at all orders of interactions, with a temporal cut-off which is dependent on the group size.
Beyond structural analysis, our library also allows to investigate a variety of dynamical processes with higher-order interactions. Here we simulate higher-order spreading among students in high school, following a model where groups of infected individuals are associated with higher-order contagion terms, in addition to traditional pairwise mechanisms~\cite{iacopini2019simplicial}. In (G) we show the fraction of infected nodes over time for two configurations, one with and one without higher-order infections. As shown, the presence of such a higher-order term might significantly change the collective dynamics, pushing the system from a healthy to an endemic state. Finally, in (H), we present a direct hypergraph visualization of the higher-order system. For simplicity, we plot individuals belonging to a single class and display all statistically significant interactions~\cite{musciotto2021detecting} among two, three and four of them.

\section*{Conclusions}
\label{sec:conclusion}

Hand in hand with new theory and methodologies, the development of efficient algorithms and software to analyze networked data has played a pivotal role in the advancement of modern network science.
Here we have presented HGX, a versatile and robust python library that offers a flexible and efficient framework to analyze networked systems with higher-order interactions. 
Its user-friendly environment and its vast range of functionalities make it accessible and useful to practitioners and researchers to answer a wide variety of needs and questions.
In the future, we aim to keep expanding the toolkit of HGX across multiple new dimensions. 
For instance, we can already foresee the implementation of tools to investigate the robustness of higher-order systems under different attack strategies. 
We will also provide methods to efficiently summarize higher-order information and reduce the dimensionality of higher-order data. 
We aim to include tools to build and analyse higher-order dependencies from multivariate time series~\cite{santoro2023higher}, and measures of information theory to capture redundant and synergistic higher-order interactions~\cite{luppi2022synergistic}. 
Moreover, we aim to expand our coverage of higher-order processes, by including different evolutionary games~\cite{alvarez2021evolutionary, civilini2023explosive}, ecological dynamics~\cite{grilli2017higher}, and more.

We hope that HGX will make higher-order network analysis open to all researchers dealing with networked data, and we invite the community to explore the library and contribute.

\section*{Acknowledgements} 
F.B. and Q.F.L. coordinated the project. Q.F.L. is the leading developer. All authors contributed tools to the library, wrote and revised the article.
F.B. and L.G. acknowledge support from the Air Force Office of Scientific Research under award number FA8655-22-1-7025. The authors also thank the International Max Planck Research School for Intelligent Systems (IMPRS-IS) for supporting M.C., the Cyber Valley Research Fund for supporting C.D.B. and M.C., the Max Planck ETH Center for Learning Systems for supporting N.R.


\bibliographystyle{ScienceAdvances}
\bibliography{bibliography}

\end{document}